\documentclass{ws-procs11x85}

\usepackage{multicol,float} %necessary when used with multicols

\begin{document}

\title{Studying the interaction between microquasar jets and their environments}

\author{M. Perucho}

\address{Max-Planck-Institut f\"ur Radioastronomie,\\
Auf dem H\"ugel, 69, Bonn 53121, Germany\\
E-mail: perucho@mpifr-bonn.mpg.de}

\author{V. Bosch-Ramon}
\address{Max Planck Institut f\"ur Kernphysik\\
Saupfercheckweg 1, Heidelberg 69117, Germany\\
E-mail: vbosch@mpi-hd.mpg.de}

\begin{abstract}
In high-mass microquasars (HMMQ), strong interactions between jets
and stellar winds at binary system scales could occur. In order to
explore this possibility, we have performed numerical
2-dimensional simulations of jets crossing the dense stellar
material to study how the jet will be affected by these
interactions. We find that the jet head generates strong shocks in
the wind. These shocks reduce the jet advance speed, and compress
and heat up jet and wind material. In addition, strong
recollimation shocks can occur where pressure balance between the
jet side and the surrounding medium is reached. All this,
altogether with jet bending, could lead to the destruction of jets
with power $<10^{36}\,\rm{erg/s}$. The conditions around the
outflow shocks would be convenient for accelerating particles up
to $\sim\,$TeV energies. These accelerated particles could emit
via synchrotron and inverse Compton (IC) scattering if they were
leptons, and via hadronic processes in case they were hadrons.
\end{abstract}


\bodymatter

\begin{multicols}{2}
\section{Introduction}\label{aba:sec1}
The interaction between a jet and its environment allows the study
of both the properties of the jet and its ambient
medium\cite{mar97,sax05}. Moreover, this interaction is important
for the evolution of jets\cite{per05}. In the case of
extragalactic jets, it is known to be different in case of FRI and
FRII sources; the collision and interaction of jets with galactic
clouds and collective stellar winds is thought to be partly
responsible for the jet mass loading and deceleration in the
former\cite{lai96}, whereas the latter can propagate and interact
with the medium up to Mpc scales. In the case of microquasars, not
much is known about the interaction of jets with their
environments\cite{hei02,bor08}, and only several sources show
evidence of such an interaction (e.g. SS~433\cite{zealey80};
XTE~J1550$-$564\cite{corbel02}; Cygnus~X-1\cite{gallo05}).

In HMMQ, the primary star suffers severe mass loss in
the form of a supersonic wind\cite{lam93}, which can embed the jet
rendering their mutual interaction likely. Here we report on a
work studying this interaction and its effects on the jet
evolution and radiation.

\section{Numerical Simulations}

We have performed three numerical simulations of jets with
cylindrical geometry and two with slab (i.e. planar\footnote{The
{\it slab jet} has the same properties as in the cylindrical case
{\bf per distance unit} in the direction perpendicular to the
simulated plane, and it is infinite in such a direction.})
geometry propagating through the wind region. We simulate the
injection of a jet, with the properties given in
Table~\ref{aba:tbl1}, in the ambient medium, i.e. the stellar
wind. The wind parameters correspond to those of a HMMQ similar to
Cygnus~X-1 or LS~5039. The density is $2.8\times 10^{-15}\,{\rm
g/cm^{-3}}$, for a star with $dM_{\rm w}/dt=10^{-6}\,{\rm
M_\odot/yr}$ at $3\times 10^{12}$~cm from the jet base. For
simplicity, at this stage the wind porosity (e.g.
Ref.~\refcite{owo06}) is assumed to be negligeable. Typically, the
velocity of an O star wind is $\sim2\times 10^8\,{\rm cm/s}$.
Despite this, the ambient is considered to be at rest in the
cylindrical simulations, this fact being, in principle, justified
because the velocity of the jet is about 100 times larger that the
wind one. It allows an axisymmetric treatment of the problem though the real
wind is moving and comes from one side, being thus intrinsically
asymmetric. We note that this assumption is not valid for weak
jets on the basis of the results of the slab geometry simulations.
The internal energy of this medium is obtained from the
temperature assumed for the wind ($10^4$ K), with a resulting
pressure of $1.5\times 10^{-3}\, {\rm erg/cm^{-3}}$. The adiabatic
index of both the ambient and the jet (supersonic and
weakly relativistic), is 5/3. The parameters in the jets are
summarized in Table~\ref{aba:tbl1}.

\begin{table}[H]
\tbl{Parameters of the simulations} {\begin{tabular}{@{}llll@{}}
\toprule Parameter & Weak j. & Mild j. & Powerful j. \\ \colrule
Jet power (erg~s$^{-1}$)& $3.0\times 10^{34}$ & $10^{36}$& $3.0\times 10^{37}$\\
Jet pressure (erg~cm$^{-3}$)& $9.1$ & $68$& $6.2\times 10^4$\\
Jet density (g~cm$^{-3}$)& $2.2\times 10^{-16}$ & $5.9\times 10^{-16}$& $1.8\times 10^{-14}$\\
Jet speed (cm~s$^{-1}$) & $1.3\times 10^{10}$ & $2.2\times
10^{10}$ & $2.2\times 10^{10}$\\ \botrule
\end{tabular}}\label{aba:tbl1}
\end{table}

The numerical simulations were performed using a two-dimensional
finite-difference code based in a high resolution shock-capturing
scheme which solves the equations of relativistic hydrodynamics.
This code is an upgrade of that described in Ref.~\refcite{mar97}.
The numerical grid is formed by 320 cells in the radial direction
and 2400 cells in the axial direction, with physical dimensions
20x300 jet radii. Extended grids were added in both directions in
order to bring the grid boundaries far from the region of
interest. The jet is injected at a distance of
$6\times10^{10}\,\rm{cm}$ from the compact object, and it is given
an initial radius of one tenth of this distance. During the
simulation, the jet covers a significant fraction of the typical
size of a HMMQ binary system (here we adopt $R_{\rm
orb}\sim3\times 10^{12} {\rm cm}$). The planar simulations were
performed with the same numerical resolution (which implies
doubling the grid radially) and a shorter axial grid by a factor
2/3.

\section{Results}

\begin{figure*}[] % always [H] in multicols
\centerline{\psfig{file=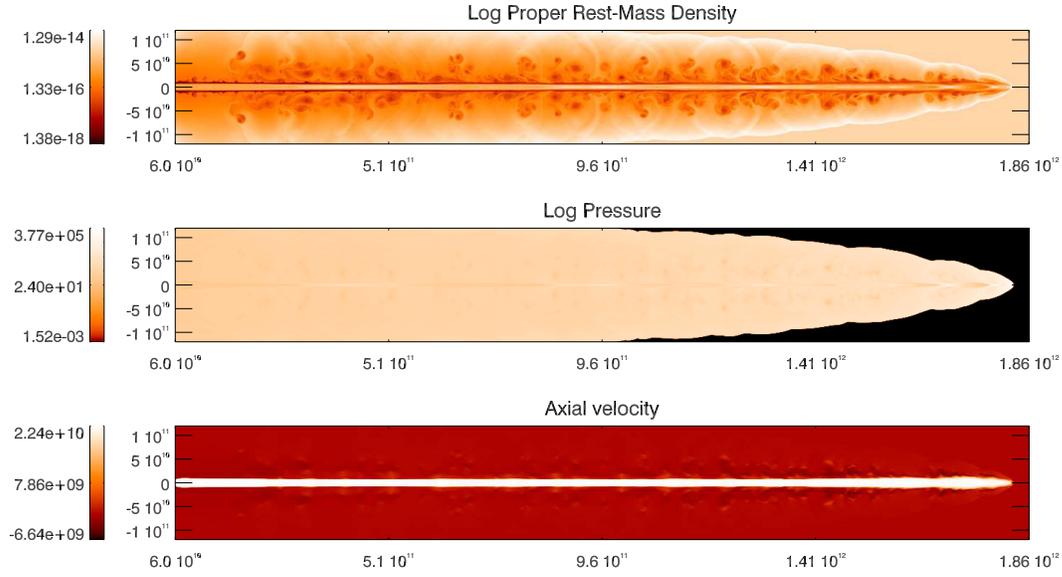,width=14cm}} \caption{Maps
of rest mass density (g~cm$^{-3}$), pressure (dyn) and axial
velocity (cm/s) at the end of the simulation ($t_{\rm
f}=164\,\rm{s}$) of the mild jet in cylindric geometry. The
horizontal and vertical coordinates indicate distances (cm) to the
compact object and the jet axis, respectively.} \label{aba:fig1}
\end{figure*}

\begin{figure*}[] % always [H] in multicols
\centerline{\psfig{file=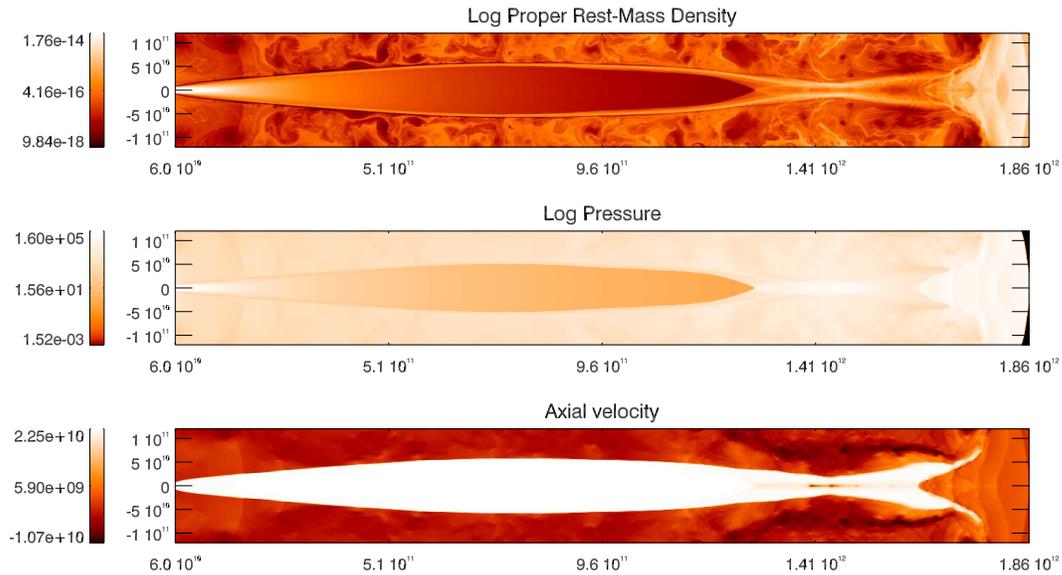,width=14cm}} \caption{Maps
of rest mass density (g~cm$^{-3}$), pressure (dyn) and axial
velocity (cm/s) at the end of the simulation ($t_{\rm
f}=212\,\rm{s}$), for a powerful jet. We have included the map of
axial velocity in order to illustrate the jump in velocity at the
recollimation shock.} \label{aba:fig2}
\end{figure*}

Figs.~\ref{aba:fig1} and \ref{aba:fig2} show the mild and powerful
jets when reaching the end of the grid, respectively. A large
pressure jump ($>10^7$) is generated at the bow shock in both
cases. A reverse shock in the jet itself is also formed. The mild
jet generates a high pressure cocoon, which keeps it collimated,
whereas the powerful jet develops a strong standing
(recollimation) shock, with a pressure jump of the order of $\sim
10^5$, due to overpressure with respect to its cocoon.
Nevertheless, the mild jet also shows weaker internal shocks with
pressure jumps of the order of $\sim 100$. The simulation of a
weak jet (not shown here) gives results which are very similar to
those of the mild jet.

Dimensional estimates\cite{per08} show that, in the jet/cocoon
phase, standing shocks must appear in the binary region ($\sim
10^{12}\,{\rm cm}$ from the jet base), at jet-temperatures $>
10^{10}\,{\rm K}$ at injection in the numerical grid. The
temperature of the mild jet is around this limit, which explains
the absence of strong shocks. However, the fact that the jet
pressure increases with decreasing distance to the compact object would
generate also a strong shock in this simulation if the injection
point of the jet in the grid were closer to the origin.

In the case of a jet interacting with a stellar wind, the lower
limit of the initial jet temperature at which a standing shock
would appear is reduced by an order of magnitude\cite{per08}. This
is important, as once the jet head is far away, the cocoon will
fade out and the stellar wind will take its place in the
interaction with the jet, though asymmetrically (only from the
direction to the primary star), what would produce asymmetric
standing shocks.

\begin{figure}[H] % always [H] in multicols
\centerline{\psfig{file=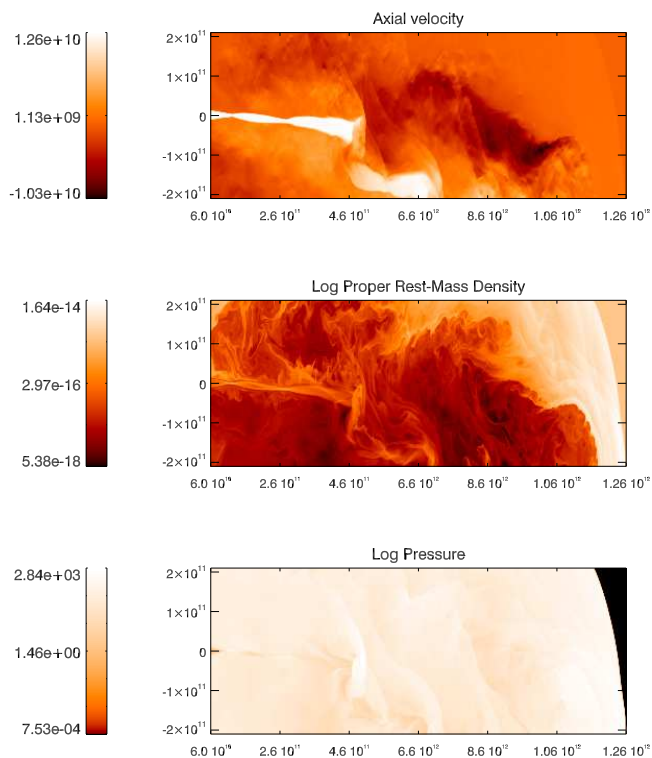,width=10cm}} \caption{Maps
of axial velocity (cm~s$^{-1}$), rest mass density (g~cm$^{-3}$)
and pressure (dyn) and axial velocity (cm/s) at the end of the
simulation ($t_{\rm f}=680\,\rm{s}$) of a weak jet in slab
geometry. The horizontal and vertical coordinates indicate
distances (cm) to the compact object and the jet axis,
respectively. The star would be on the vertical axis of the plot,
left of the jet, at the orbital distance (out of the plot).}
\label{aba:fig3}
\end{figure}

We have also performed planar jet simulations for the weak and
mild jet, in which the wind velocity is taken into account.
Fig.~\ref{aba:fig3} shows the influence of a lateral wind on the
weak jet. In these maps, the asymmetrical standing shocks produced
by the jet-wind interaction are clearly observed. A second
simulation for a slab jet with the same physical parameters as the
mild cylindrical jet shows that a jet with power $10^{36}\,{\rm
erg/s}$ can also be destroyed by such a wind\cite{per08}.
3-dimensional simulations will be performed in order to check this
interaction in a more realistic setup.

\section{Non-thermal radiation from jet-wind interactions}

It is worthy noting the importance of the strong shocks predicted
by our simulations from the radiative point of view. On one hand,
particle acceleration is likely to take place in the standing
shocks or the head of the jet when crossing the denser wind
regions. The presence of a non thermal population of electrons
plus the strong photon field produced by the primary star will
lead to strong IC emission (e.g. Ref.~\refcite{kan07,per08}).
Wind-jet interactions may be the mechanism triggering particle
acceleration in several massive X-ray binaries with jets from
which gamma-rays have been detected\cite{aha05,alb06,alb07}. In
addition, these shocks will possibly increase also the value of
the magnetic field (not considered in the simulation at this
stage) in the postshocked regions to high values, allowing for
efficient synchrotron radiation up to X-rays and beyond (e.g.
Ref.~\refcite{per08}). Moreover, when enough relativistic proton
power were available, there may be significant gamma-ray
production via decay of neutral pions from proton-proton
interactions due to shock enhancement of the matter density, plus
other products like secondary pairs and even neutrinos (e.g.
Ref.~\refcite{rom03,aha06}). We note that for a 10\% efficiency
for energy transferring from shocks to non thermal particles,
multiwavelength leptonic luminosities of $\sim10^{36}\,{\rm erg/s}$,
detectable from low (e.g. radio) to very-high energies (e.g. TeV),
may be easily reached in the case of a powerful jet.

\section{Summary}

We show that, with the reasonable overpressure and physical
parameters in the jets and stellar winds of HMMQ, assuming
negligible wind porosities, the jets form strong standing shocks
due to interaction with the cocoons formed first, and the stellar
wind from the primary later, on top of the transient reverse/bow
shocks. We estimate that standing shocks must form within the
system region for typical jet temperatures ($>10^{9-10} {\rm K}$).
Moreover, we have shown that jets with power $<
10^{36}\,\rm{erg/s}$ could be deflected and disrupted by the primary
star wind. The influence of hydrodynamical instabilities such as
Kelvin-Helmholtz is to be neglected on the basis of our results,
as the evolution and dynamics of these jets are nonlinear from the
start. In this scenario, the inertia of the jet compared to that
of the wind is responsible for the fate of the former. Finally,
the interaction of the jet head with the wind, and the jet sides
with the cocoon and the stellar wind, in HMMQ, could lead
to efficient non-thermal multiwavelength radiation, showing the
importance of the role of the primary star for high energy
emission in these sources, which is not just a provider of seed
photons for IC scattering and photon photon absorption,
or target nuclei for hadronic processes.

\section*{Acknowledgments}

\end{multicols}

\end{document}